\documentclass{optica-article}

\journal{opticajournal} 

\articletype{Research Article}

\newcommand{\cref}[1]{Chapter~\ref{#1}}  

\usepackage{graphicx} 
\usepackage{cite}
\usepackage{lineno}
\usepackage{gensymb}
\usepackage{float}

\begin{document}

\title{Galvanometer-scanning transient phase microscopy with balanced detection and arbitrary pump polarization}

\author{Cameron N. Coleal,\authormark{1} Randy A. Bartels,\authormark{2} and Jesse W. Wilson\authormark{3,4*}}

\address{
\authormark{1}Los Alamos National Laboratories
\authormark{2}Morgridge Institute for Research
\authormark{3}School of Biomedical and Chemical Engineering, Colorado State University
\authormark{4}Department of Electrical and Computer Engineering, Colorado State University
}

\email{\authormark{*}Corresponding author: jesse.wilson@colostate.edu} 


\begin{abstract*} 

Transient absorption microscopy measures excited-state kinetics based on the imaginary part of the pump-induced perturbation to the complex refractive index, i.e. $\Im \{\Delta\mathcal N\}$, with applications in both materials and biomedical sciences. Its complement, transient phase microscopy, enabled by stable inline birefringent interferometry, measures the real part $\Re\{\Delta \mathcal N\}$. The ability to switch between absorption and phase measurements may yield a stronger signal, depending on the sample and probe wavelength. To date, however, transient phase has not been coupled with galvanometer scanners, thus limiting it to materials science applications and non-imaging spectroscopy. Here, we extend transient phase microscopy to operate in a galvanometer-scanning microscope with balanced detection, comparing amplitude and phase measurements in graphene (in which amplitude detection has the advantage), hemoglobin and red blood cells (in which phase detection has the advantage). We examine the impacts and limitations introduced by galvanometer scanning, in addition to relocation of the pump-probe combining dichroic to permit arbitrary polarization of the pump.

\end{abstract*}

\section{Introduction}

Pump-probe transient absorption spectroscopy is a valuable tool for probing femtosecond chemical reactions and excited-state molecular relaxation pathways \cite{Zewail_1988}. When coupled with a laser-scanning microscope, transient absorption microscopy (TAM) \cite{Fischer_Wilson_Robles_Warren_2016} generates contrast from absorptive materials through pump-induced changes in the complex refractive index $\Delta \mathcal N$ (e.g. excited-state absorption, ground state depletion, stimulated emission, transient changes in Franck-Condon broadening, and transient photoproducts). TAM can also measure, at each pixel, lifetimes associated with relaxation pathways and kinetics (e.g. spontaneous emission, vibrational cooling, and photodissociated ligand recombination). The differences in excited-state lifetimes can provide contrast between materials that have very similar absorption spectra, such as melanins \cite{MelaninPP_1_Warren_2015,MelaninPP_2_Warren_2016,MelaninPP_3_Warren_2019}. Other applications include hemoglobin spectroscopy and imaging \cite{HemoglobinPP_1_Fu_2019,HemoglobinPP_2_Fu_2020} and emerging techniques to evaluate mitochondrial redox through cytochrome electron transport hemeproteins \cite{ErkangPP_1_Wilson_2022,ErkanPP_2_Wilson_2022}, along with numerous applications in materials sciences \cite{MaterialSciApp_1_Li_2021,MaterialSciApp_2_Wang_2019, MaterialSciApp_3_Ma_2022, ExtractingQuantitativeDielectricFxn_2022} and cultural heritage \cite{HeritagePP_1_Fischer_2014, HeritagePP_2_Ji_2023}. With few exceptions, TAM operates by measuring pump-induced changes to the detected amplitude of a probe pulse, following excitation after a delay of $\tau$, i.e. $S_{\rm TAM}\propto \Im\{\Delta \mathcal N (\tau)\}$, where $\Im$ denotes the imaginary part of a complex quantity. In these experiments, images are formed by raster scanning the combined, focused, pump and probe beams across the sample. This can be done either by mechanically translating the sample (typical for materials science applications), or by imaging a pair of galvanometer-mounted mirrors to the back aperture of a microscope objective to scan the beam (typical for biomedical applications). Galvanometer scanning is more complex, but has the advantage of speed, enabling the capture of dynamics and reducing sample heating by reducing pixel dwell time.

The complementary measurement by transient \emph{phase} microscopy (T$\Phi$M), senses $S_{\rm T\Phi M}\propto \Re\{\Delta \mathcal N(\tau)\}$, where $\Re$ denotes the real part, through interference between the probe and a reference pulse, although diffraction, scattering, or lensing could also be exploited \cite{bartels2021low}. In pump-probe spectroscopy, such phase measurements are most often associated with impulsive Raman scattering (ISRS), where molecular bond deformations or rotations are directly coupled to $\Re\{\Delta \mathcal N(\tau)\}$ via the differential polarizability \cite{bartels2001phase, bartels2002nonresonant, Wilson_JosaB_THGVibSpec_2012, bartels2021low} and can be measured through interferometry \cite{hartinger2008single, Wilson_SyntheticAperture_2008, wilson2008phase, schlup2009sensitive, MichelsonInterferometer_Wahlstrand_2023} or spectral shifting \cite{domingue2014time, Smith_IRS_Bartels_2021, smith2022nearly}. A similar situation is found with pump-induced molecular rotational motion \cite{Robles_PPNLDS_Warren_2013}. On the other hand, in pump-probe imaging and spectroscopy of absorptive pigments, $\Re\{\Delta \mathcal N(\tau)\}$ is largely neglected, owing to the simplicity of measuring $\Im\{\Delta \mathcal N(\tau)\}$, which does not require stable probe-reference interference.

While similar in their ability to measure transient kinetics, there may be some scenarios where T$\Phi$M has advantages over TAM. For example, consider the ground state bleaching response of an absorption band. After pumping to the excited state, the maximal change in absorptivity will occur at the peak wavelength of the ground state absorption band. Thus, for an absorption measurement, the probe wavelength yielding the strongest response will be at the absorption peak, and will thus achieve poor penetration into thick materials. For a phase measurement, however, the maximum change in refractive index response inferred from Kramers-Kronig relations will be offset from the absorption peak. Thus, the probe wavelength yielding the strongest phase response will not suffer the same depth limitations.

Various techniques have been developed to measure $\Re\{\Delta\mathcal N\}$ through its modulation of probe phase in limited imaging applications measuring only the instantaneous (non-transient) cross-phase modulation (XPM) part of $\Re\{\Delta\mathcal N\}$) and in non-imaging applications. The simplest measurement detects spectral shifting proportional to the time derivative of the modulation, as has been done in XPM imaging of biological samples~\cite{XPMSS_wilson} and is frequently employed in ISRS spectroscopy and imaging \cite{bartels2021low, Smith_IRS_Bartels_2021}. A more direct measurement of probe phase can be done by interference with a reference pulse. Early work in ISRS focused on spectral interferometry, where the time-delayed probe and reference pulses produce interference fringes on a spectrometer \cite{SagnacInterferometer_Kobayashi_1995, MichelsonInterferometer_Wahlstrand_2023}. The fringe spacing is inversely proportional to probe-reference delay, and pump-induced $\Re\{\Delta \mathcal N\}$ causes the fringes to shift. Alternatively, probe and reference pulses can be re-timed and overlapped on a single-element detector \cite{van2005detection, Orrit, Guillet_2015}. Such measurements are spoiled by any relative phase instability between probe and reference, and require geometries where probe and reference trace a common path from the point where they are split from a common initial pulse to the point where their interference is measured after interaction with the sample. The challenge in common-path interferometry is to ensure that the probe interacts with the pump-induced perturbations to the sample, while the reference does not. This can be done in a Sagnac interferometer in which counter-propagating probe and reference pulses arrive at different times in the sample~\cite{SagnacInterferometer_Kobayashi_1995,Theisen2025}, or by propagating probe and reference along slightly different paths so that the pump overlaps spatially only with the probe~\cite{MichelsonInterferometer_Wahlstrand_2023}, or by using group delay differences in bulk birefringent media to create a probe-reference pair, which arrive at different times in the sample~\cite{Schlup}. (It should be noted that for short probe-reference delays, these methods measure a finite difference $\Re\{\Delta\mathcal N(\tau_{\rm pr})-\Delta\mathcal N(\tau_{\rm ref})\}$ rather than $\Re\{\Delta\mathcal N(\tau_{\rm pr})\}$ directly \cite{Wilson_SyntheticAperture_2008, bartels2021low}.)

Of these, the birefringent common path interferometer provides a simple means of re-timing probe and reference, enabling measurement with a single-element detector rather than a spectrometer, and has been employed for non-imaging spectroscopy. In a transmission setup with a pair of pulse-splitting and re-timing calcites~\cite{Orrit}, measured both the real and imaginary components of the perturbed optical transmission coefficient for gold nanoparticles. Similarly, ~\cite{Guillet_2015} made complex measurements of the perturbed optical reflection coefficient for a tungsten thin film. In their setup, the probe and reference are split on their way to the sample by a birefringent crystal, then after reflection from the sample, re-timed by propagating in the reverse direction through the same crystal. A half-wave plate can shift the measurement from phase to amplitude in both studies, permitting measurement of the complex $\Delta\mathcal N$. This approach to T$\Phi$M has yet to be implemented on a galvanometer-scanning microscope, and thus has not been applied to imaging of biological samples. In addition, the placement of the pump-combining dichroic before the birefringent splitter precludes rotation of pump polarization angle, which can have a large effect on time-resolved spectroscopy of biological pigments such as melanin \cite{Grass_PolarizationMelaninPP_Warren_2022}.

Here, we extend T$\Phi$M with birefringent inline interferometry to a galvanometer-scanning microscope, imaging biological and crystalline samples in a transmission geometry. With this setup we are able to compare two primary differences between TAM and T$\Phi$M: (1) differences in the SNR for a particular sample at our set pump and probing wavelengths and (2) additional sample information revealed by instantaneous phase interactions (i.e. cross-phase modulation, XPM) versus instantaneous absorption interactions (i.e. two-photon absorption, TPA), which is particularly revealing in an imaging geometry. In contrast to previous works ~\cite{Orrit,Guillet_2015}, we place the pump-combining dichroic after the probe-reference birefringent splitter, allowing for the freedom to arbitrarily rotate the pump polarization. We evaluate the impact that galvanometer scanning has on probe-reference re-timing and field of view, in addition to the signal-to-noise impact of polarization effects from placing the dichroic after the birefringent probe-reference splitter~\cite{DichroicAberrations}. Furthermore, we implement a balanced detection scheme that doubles the T$\Phi$M signal while canceling relative intensity noise from the laser. We demonstrate spectroscopy and imaging measurements on graphene (in which TAM yields a stronger signal) and hemoglobin and red blood cells (in which T$\Phi$M yields a stronger signal). Our results demonstrate the feasibility of galvanometer-scan imaging using transient phase detection and suggest that the ability to switch between TAM and T$\Phi$M can yield an improved SNR for a given set of pump and probe wavelengths.


\section{Methods}

\subsection{Experimental Setup} 

A Yb-doped fiber laser oscillator (Menlo YLMO) seeds a fiber chirped pulse amplifier (KMLabs) that provides pulses at $f_{\rm rep}=5~{\rm MHz}$ centered at a fixed $\lambda=1035~{\rm nm}$ with a bandwidth of $\sim8.25$ nm. This output is split into a pump pulse train for the experiments, and another portion feeds a  noncollinear optical parametric amplifier (KMLabs) to generate the probe and reference pulses at $\lambda=800~{\rm nm}$ with a bandwidth of $\sim12$ nm. Afterwards, dispersion compensation is managed on the pump arm with a grating compressor and a prism compressor, and on the probe arm with a prism compressor. We used the fixed output for the pump and the tunable output for the probe to enable future studies on probe wavelength dependence. 

\begin{figure*}[htbp]
    \centering
    \includegraphics[width=0.85\textwidth]{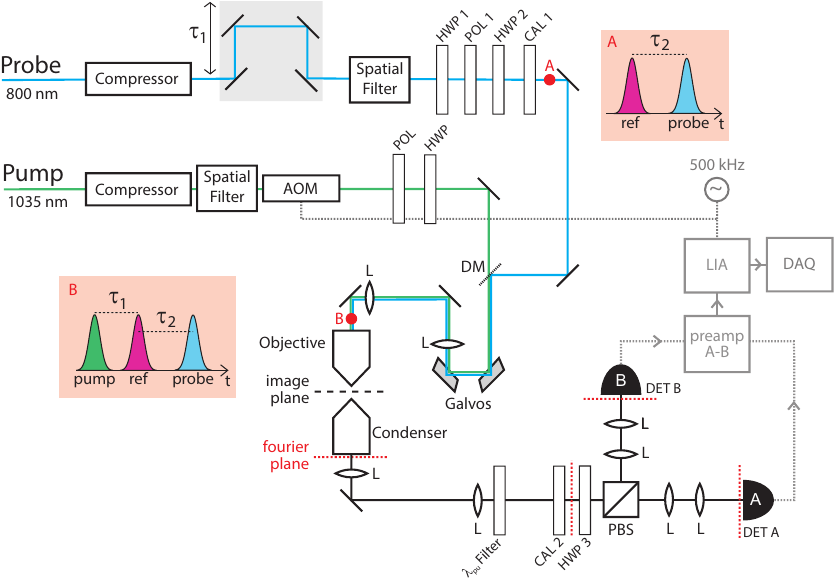}

    \caption{An overview of the system. Prism and grating compressors are utilized to compress the probe and pump pulses by accounting for dispersion accumulated throughout the beam line. Spatial filters utilizing a 4-f conjugated setup with a pinhole at the Fourier plane are used to clean up the spatial modes of both beams. The probe pulse is split into a reference and probe pulse with the first calcite crystal (CAL 1) separated by a time delay of $\sim$1.2 ps (see Label A). The delay stage in the probe beam controls the temporal separation of the pump and reference pulses (see Label B). We utilize a Nikon Plan-Apo $\lambda$ 20X objective (0.75 NA) and a Zeiss LD Epiplan 50X condenser (0.50 NA). The Fourier plane on the back aperture of the condenser is relayed to the plane between the re-timing calcite (CAL 2) and half-wave plate (HWP 3), as well as to the detector planes. (Galvos: galvanometer scan mirrors, HWP: half-wave plate, POL: polarizer, CAL: calcite, DM: dichroic mirror, L: lens, $\lambda_{\rm pu}$ Filter: pump rejection filter, DET: detector, LIA: lock-in amplifier, DAQ: data acquisition device) }
    \label{fig:SystemDiagram}
\end{figure*}

The system diagram after the laser source is shown in Fig.~\ref{fig:SystemDiagram}. A motorized stage delays the probe beam (prior to splitting into a reference-probe pulse-pair) relative to the pump by $\tau_1$. (After the probe is split by CAL 1 into a reference-probe pulse-pair this temporal separation, $\tau_1$ is defined between the pump and the leading pulse from the pulse-pair, i.e. the reference pulse, inset B). A set of polarization optics (HWP 1, POL 1, HWP 2) cleans up the probe polarization and rotates it to $45^\degree$ with respect to the eigenaxes of a 2~mm thick calcite (CAL 1), with its optic axis oriented in the plane of the table. This results in an equal projection of the probe pulse along the fast (extraordinary) and slow (ordinary) axes of the crystal, respectively, resulting in two output pulses. The p-polarized reference exits first and is  followed $\tau_2\sim1.2~{\rm ps}$ later by an s-polarized probe (see Supplementary 1A). The pump acquires a 500 kHz sinusoidal amplitude modulation by an acousto-optic modulator (AOM; TEM-100-10-1050, Brimrose Corp., Baltimore, MD). 
A polarizer and half-wave plate provide rotatable pump polarization, before combining with the probe-reference pair in the dichroic mirror (DM) (Thorlabs DMLP900). The combined pulses are directed into a laser scanning microscope, with a pair of $x/y$ galvanometer scan mirrors (QS-7, Nutfield Technology, Inc., Windham, NH) conjugated to the back focal plane of a microscope objective, followed by the sample, another objective (the condenser), and a detection arm. The objective has a numerical aperture (NA) of 0.75, leading to an estimated $1/e^2$ diameter of $880~{\rm nm}$ for the pump, $680~{\rm nm}$ for the probe, and $540~{\rm nm}$ for the product of the two beams at the focus (neglecting aberrations and assuming Gaussian profiles).

A pump reject filter eliminates the pump pulse train after the condenser and the reference-probe pair is relayed from the condenser to the re-timing calcite (CAL 2). CAL 2 is oriented so as to undo the relative delay $\tau_2$ between the probe and reference, with its slow (ordinary) axis aligned with the advanced reference pulse and fast (extraordinary) axis aligned with the delayed probe pulse. Afterwards, probe and reference are temporally overlapped and in quadrature ($\pi/2$ relative offset phase, adjusted by fine-tuning the azimuthal angle of the splitter calcite) but orthogonally polarized. A half-wave plate then rotates probe and reference so each is $\pm 45^\degree$ with respect to the axis of a polarizing beam splitter (PBS). Detector A (Thorlabs PDA36A2) measures the p-polarized projection and Detector B (Thorlabs PDA36A2) the s-polarized projection. In this configuration, an advance in probe phase relative to the reference increases A while decreasing B. Rotating the half-wave plate can reconfigured for TAM, rather than T$\Phi$M, as detailed below. Throughout, relay lenses are used to image the Fourier plane (collection objective back aperture) to the re-timing calcite and the detectors, minimizing the effects of lateral movement during image scanning applications. 

\subsection{Detection Technique} 

\begin{figure}[H]
    \centering
    \includegraphics[scale = 0.70]{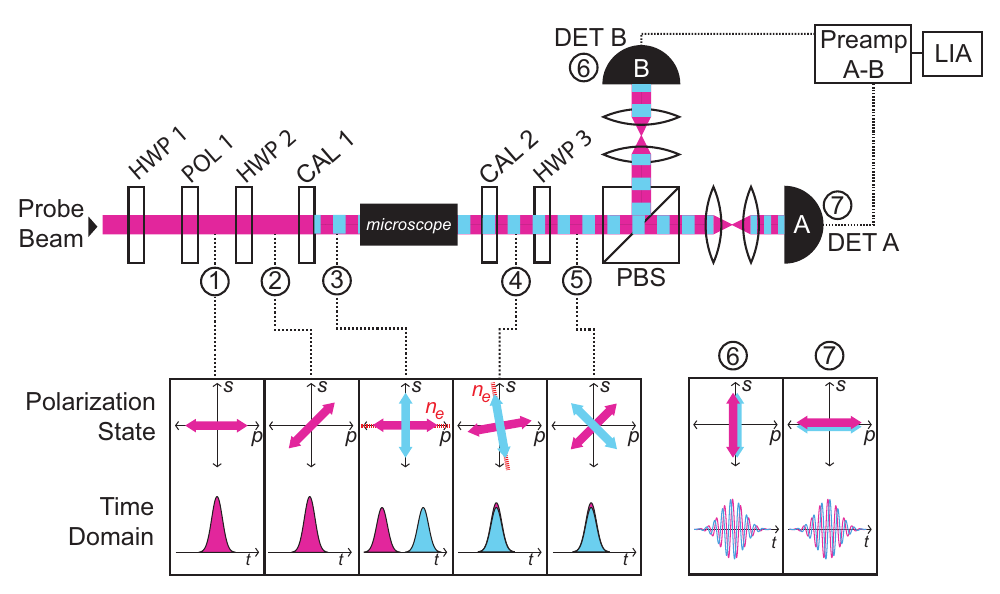}
    \caption{Overview of the polarization and detection optics and their impact on the reference and probe pulses.}
    \label{fig:DetectionOptics}
\end{figure}

A diagram focusing on the polarization optics is shown in Figure \ref{fig:DetectionOptics}. The probe proceeds first through a half-wave plate (HWP 1) tuned to maximize the amount of light that proceeds through the polarizer (POL 1), resulting in linear p-polarization (parallel to the plane of the table) at step 1. HWP 2 rotates the polarization by $45^\circ$ (step 2). CAL 1 is a 2 mm thick piece of a-cut calcite (optic axis runs along the front face). It is oriented so that p-polarization sees the extraordinary axis and s-polarization sees the ordinary axis. Thus the probe is projected onto the ordinary ($n_o = 1.6488, n_{g,o} = 1.6737$ \cite{Ghosh_1999_RIValues}) and extraordinary axes ($n_e = 1.4819, n_{g,e} = 1.4920$ \cite{Ghosh_1999_RIValues}). Here $n_i$ and $n_{g,i}$ for $i \in {o,e}$ are the refractive index and group index, $n_{g,i} = (n_i - \omega \, \partial_\omega \, n_i)_{\omega = \omega_{\rm pr}}$ evaluated at the center frequency of the probe pulse, $\omega_{\rm pr}$. Two pulses thus exit CAL 1 (step 3) orthogonally polarized; the reference is advanced and p-polarized, the probe is delayed and s-polarized.  The pair proceeds through the microscope before encountering CAL 2. CAL 2 is oriented opposite to CAL 1; the ordinary (slow) axis is rotated to coincide with the advanced reference pulse and the extraordinary (fast) axis coincides with the delayed probe pulse. The result is a re-timing of the two pulses (step 4). Note that the axes of CAL 2 may not align with s- and p-polarization as referenced to the table surface, but is instead based on the axes of the reference and probe polarizations after going through the microscope (this is due to the orientation of the detection arm which utilizes a vertical beam path relative to the tabletop). HWP 3 is used to rotate the re-timed reference and probe polarizations such that they are 45 degrees to the s- and p- axes of the polarizing beam splitter (PBS), so that both the reference and probe pulses arrive at both DET A and DET B (step 5). The pulse pair on DET B (step 6) will be $\pi$ out of phase from the pulse pair on DET A (step 7). Thus, if one detector measures destructive interference between the reference-probe pair then the other detector will measure constructive interference.

\begin{figure}
    \centering
    \includegraphics[scale = 0.65]{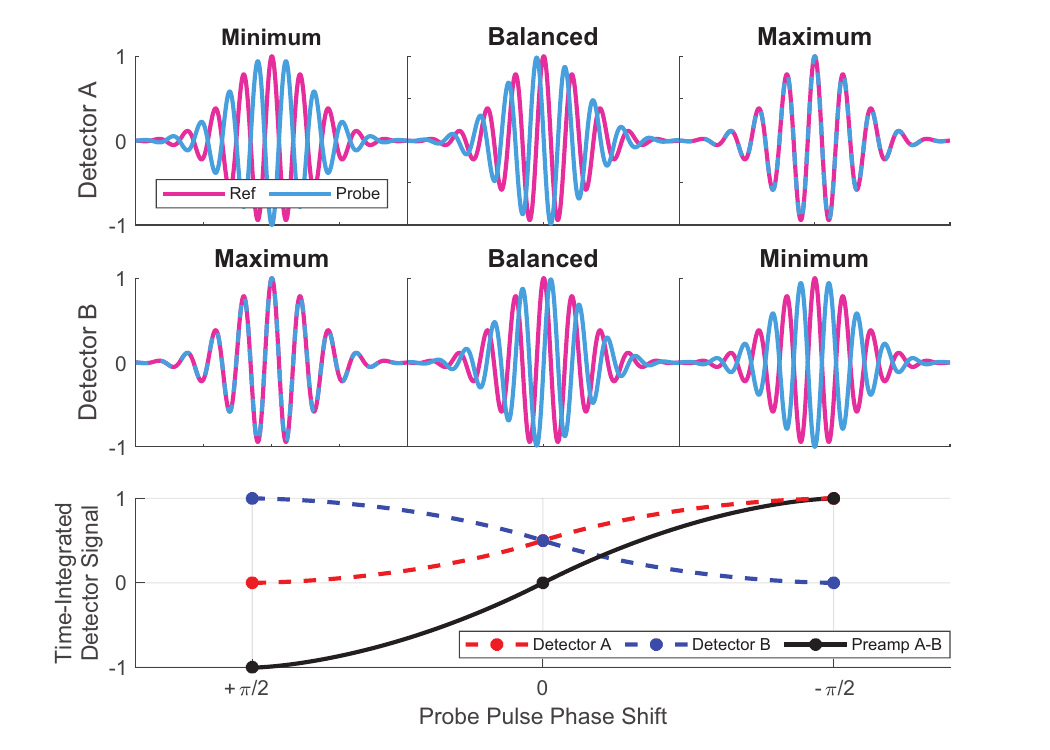}
    \caption{Visual of the interference signals present on Detector A and B with different phase shifts imparted to the probe pulse.}
    \label{fig:PreampSignal}
\end{figure}

With each sample, the initial offset phase ($\Phi_{\rm offs}$) between the probe and reference is set to a balanced position such that $V_A - V_B = 0$, where $V_A$ is the voltage on DET A and $V_B$ is the voltage on DET B. The offset phase can be set by adjusting the angle-of-incidence ($\alpha_{\rm ext}$) of CAL 1 with p-polarized light. This angle is set with the pump pulse blocked so that the offset phase accounts for any residual phase imparted to the reference and probe pulses by the microscope and sample. When the pump pulse is unblocked, a phase advance imparted to the probe pulse from an excited sample will cause an increase in $V_A$ and a decrease in $V_B$ and the overall balanced detection signal $V_A - V_B$ will swing positive. Similarly, a phase delay imparted to the probe pulse will cause a decrease in $V_A$ and an increase in $V_B$ and the overall signal $V_A - V_B$ will swing negative. Figure \ref{fig:PreampSignal} shows an example of the temporal interference between the reference (pink) and probe (blue) pulses on detector A and B for these scenarios. 

Ultimately our setup is capable of measuring an excited-state phase change of $\pm\frac{\pi}{2}$ from the initial offset point. The voltage measured on each detector is the time-integrated interference between the reference and probe; since our final signal is $V_A - V_B$ we end up with a 2X increase in measurable signal than if we were conducting our measurements with a single detector. (It should be noted, though, than both TAM and T$\Phi$M can be done with only one detector.) The preamplifier (SR 560) measures $V_A - V_B$ and applies a voltage gain before the difference signal is sent into a lock-in amplifier (Liquid Instruments Moku:Lab). The lock-in amplifier measures amplitude changes of $V_A - V_B$ that coincide with the 500 kHz pump modulation frequency; these measurements correspond to pump-induced phase changes $\Re\{\Delta \mathcal N\}$.

The balanced detection signal is given by
\begin{equation}
    \Delta V \propto \cos( \Delta\Phi_{\rm t\phi m} + \Phi_{\rm offs} ).
\end{equation}
where $\Phi_{\rm offs}$ is a phase offset between reference and probe set by CAL1, CAL2, any birefringent optics in the path, e.g. the pump-combining dichroic, and static birefringence of the sample (see Supplementary 2 and 3 for more details). The term arising from pump-probe interactions is $\Delta\Phi_{\rm t\phi m}\propto\Re\{\Delta \mathcal N\}$. In the balanced configuration, $\Phi_{\rm offs}=0$, and assuming that the phase perturbations resulting from $\Re\{\Delta \mathcal N\}$ are small,
\begin{equation}
    \Delta V \propto \Re\{\Delta \mathcal N\}.
\end{equation}

\begin{figure}
    \centering
    \includegraphics[scale=0.75]{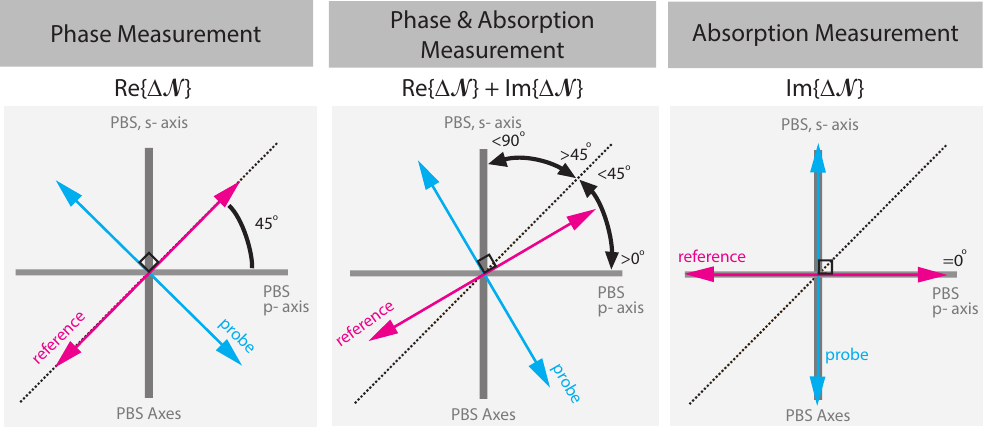}
    \caption{Polarization angle with respect to PBS axes determines if the measurement will be a phase measurement, an absorption measurement, or a combination of the two.  }
    \label{fig:MeasurementType_HWPangle}
\end{figure}

One advantage of this detection configuration is the ability to switch between measuring transient absorption ($\Im\{\Delta \mathcal N\}$) and transient phase ($\Re\{\Delta \mathcal N\}$) simply by changing the projection of the probe and reference pulses on the PBS axes \cite{Orrit,Guillet_2015}. This projection angle is controlled by HWP 3; Figure \ref{fig:MeasurementType_HWPangle} shows three projection scenarios of the reference-probe pair polarization axes onto the PBS. If HWP 3 is oriented such that the reference-probe pair polarization axes are 45 degrees to the s- and p- axes of the PBS (left image), then there is an equal amplitude projection of both pulses to the two detectors. This setting is for making purely a \textit{phase} measurement since any changes to the amplitudes of either pulse will cancel in the balanced detection scheme; only changes to the temporal interference on either detector caused by changes in $\Re\{\Delta \mathcal N\}$ will result in $V_A - V_B \neq 0$.

If HWP 3 is oriented such that the reference-probe pair polarization axes are aligned to the s- and p- axes of the PBS (right image), then DET A will measure amplitude changes to the reference pulse and DET B will measure amplitude changes to the probe pulse. This setting is for making purely an \textit{absorption} measurement since any changes to the phase of either pulse will have no bearing on the integrated intensity measured by the detector; only changes to the amplitude of either pulse caused by changes in $\Im\{\Delta \mathcal N\}$ will result in $V_A - V_B \neq 0$.

Finally, if HWP 3 is oriented for any other projection (middle image), then a combination of a phase and absorption measurement will be made.

\section{Results}

\subsection{Transient Phase and Absorption Measurements}

This section presents the results of measuring transient absorption and transient phase in glass, a graphene monolayer, bovine hemoglobin in solution and fresh human red blood cells.

\subsubsection{Cross-Phase Modulation Measurement in Glass}

\begin{figure}
        \centering
    \includegraphics[scale = 0.8]{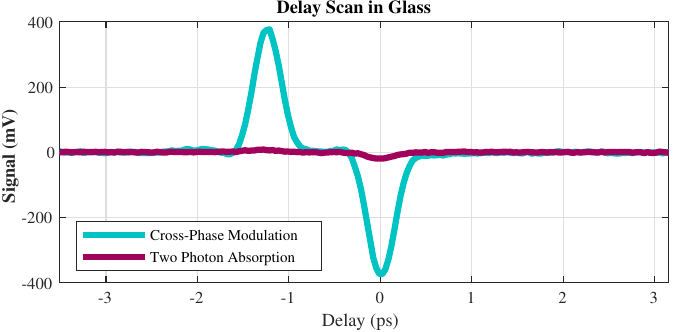}
    
    \caption{An initial delay scan in borosilicate glass showing cross-phase modulation detectable by transient phase, and a very small two photon signal measured with transient absorption. $3.9$ mW pump and $0.78$ mW probe incident on the sample.}
    \label{fig:delayScanGlass}
\end{figure}

First we acquired a delay scan in a glass microscope slide (Figure \ref{fig:delayScanGlass}). Glass has a minimal absorption at the probe wavelength of 800 nm and the interaction that occurs between the sample and the pump pulse is instantaneous. When making an absorption measurement, we would expect to see a two-photon absorption peak when the pump and probe pulses were overlapped if the sample absorbed the pump wavelength. For a phase measurement, we are able to see cross-phase modulation (XPM) from the instantaneous optical Kerr effect, in which pump intensity changes the refractive index seen by the probe~\cite{Wilson_SyntheticAperture_2008}. Consequently, due to the instantaneous nature of the interaction, the width of the XPM pulses indicates a $\sim 350~{\rm fs}$ cross-correlation between pump and probe. (Frequency resolved optical gating (FROG) measurements were previously made that showed the probe pulse width was the limiting factor).

\subsubsection{Graphene Monolayer} 

Next we examined a graphene monolayer on a 0.5 mm quartz substrate placed on top of a 1 mm glass microscope slide. Previous works looking at graphene have utilized transient absorption to analyze graphene carrier dynamics \cite{GrapheneTAS_SingleLayerHyperspectral_Gesuele_2019, MaterialSciApp_3_Ma_2022}. Near-infrared pumping (as in our setup) results in transient heating of free carriers~\cite{Wagner2014}. It is a material that shows promise for use in various applications including electronics and drug delivery, making it a popular research target \cite{Graphene_TAM_InBlood_Yang_2015}.

\paragraph{Time-Domain Spectroscopy}

\begin{figure}
    \centering
    \includegraphics[scale = 0.75]{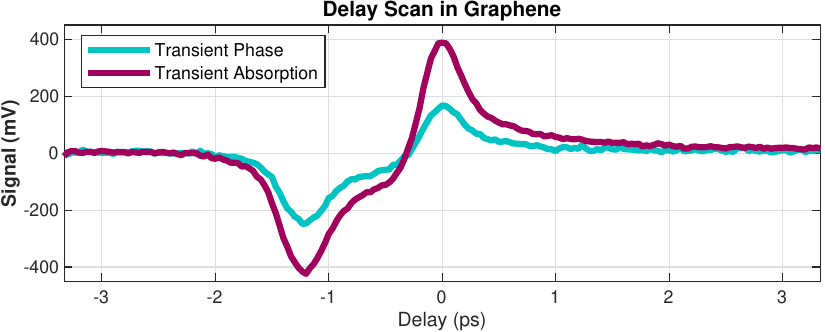}
    \caption{A transient absorption versus transient phase delay scan in monolayer graphene. At the focus, the pump power is $7.8$ mW and the probe power is $0.97$ mW.}
    \label{fig:GrapeneDelayScan}
\end{figure}

Figure~\ref{fig:GrapeneDelayScan} shows a delay scan taken in the center of the sample.  A stronger transient absorption versus transient phase response is observed, with a decay time of $\sim$2 ps (found by fitting to an exponential function). The first peak at $\tau = -1.2$ ps corresponds to the pump-reference overlap and the second peak at $\tau = 0$ ps corresponds to the pump-probe overlap. 

\begin{figure}
    \centering
    \includegraphics[scale = 0.7]{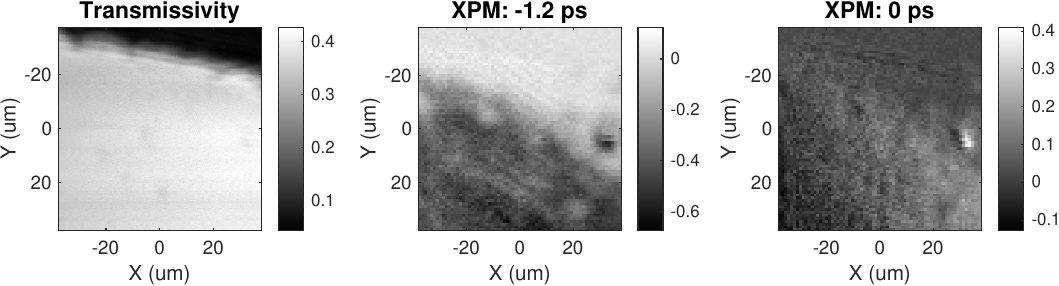}
    \caption{Graphene transmissivity image and the peak XPM images.}
    \label{fig:GrapheneTransmission}
\end{figure}

\begin{figure}
    \centering
    \includegraphics[scale = 0.83]{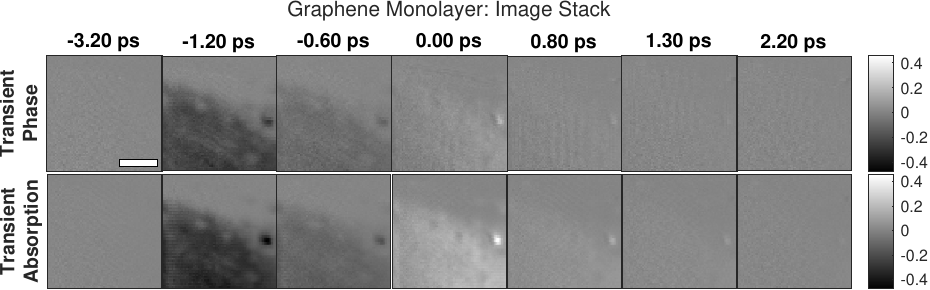}
    \caption{Delay stacks for graphene with (top row) T$\Phi$M and (bottom row) TAM. Intensity measured in voltage from the lock-in amplifier output. The scale bar is 25 $\mu$m wide. At the focus, the pump power is 7.8 mW and the probe power is 0.97 mW. }
    \label{fig:Graphene_ImageStack}
\end{figure}

\paragraph{Imaging} 
Image stacks were acquired at the edge of the graphene sample where spatially-varying structure is expected. Images of $100\times100$ pixels were acquired with 0.05 second pixel dwell time (8.3 minutes/frame). A probe transmissivity image was acquired after setting the half-waveplate for phase measurements and the angle of incidence for the first calcite for a "balanced" state (i.e. $V_A-V_B=0$) with the pump blocked. From Figure \ref{fig:GrapheneTransmission} the edge of the sample can be seen in this transmissivity image (left-most image) and gives context to interpreting the image stack in Figure \ref{fig:Graphene_ImageStack}. Within the transmissivity image the edge of the sample is seen at the transition from light to dark (where the light pixels correspond to the sample)

An image stack was acquired for both transient phase and transient absorption at the edge of the graphene and are compared (Fig.~\ref{fig:Graphene_ImageStack}). For this sample, we see that transient absorption provides a stronger signal at overlap with both the probe and reference (-1.20 ps and 0.00 ps) than transient phase. 

\subsubsection{Hemoglobin and Red Blood Cells}

Next we moved onto biological samples. First, we looked at the spectroscopy of a hemoglobin sample prepared with 9 mM of bovine hemoglobin lyophilized powder (Sigma-Aldrich H2500) in phosphate buffer solution.

Second, we imaged a droplet of whole blood placed on a poly-L-lysine coated microscope slide with a glass coverslip. The whole blood used was acquired from healthy donors by venipuncture by Dr. Liszt Madruga of the Department of Chemical Engineering (CSU) working under Prof. Matt Kipper of the Departments of Chemical \& Biological Engineering (CSU) and Prof. Ketul Popat of the Department of Mechanical Engineering (CSU). The protocol followed was approved by the Colorado State University Institutional Review Board. 

\paragraph{Time-Domain Spectroscopy of Hemoglobin} 

\begin{figure}
    \centering
    \includegraphics[scale = 0.75]{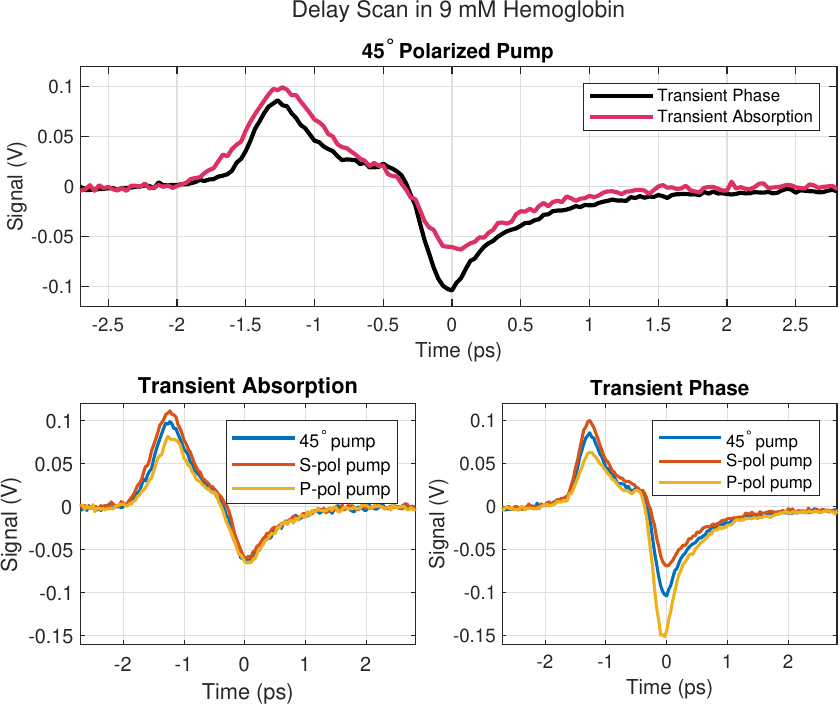}
    \caption{Delay scans of hemoglobin in solution, with 0.81 mW pump and 0.73 mW probe at the sample. The top row shows a comparison (in V) of the measured transient phase (black) and transient absorption (pink) delay scans for a $45\degree$ polarized pump beam. The first peak occurs with overlap of the pump-reference pulses at $\tau=-1.2$ ps and the second peak occurs with overlap of the pump-probe pulses at $\tau=0$ ps. The bottom left image compares three transient absorption delay scans for different pump polarizations and the bottom right image compares three transient phase delay scans for different pump polarizations.}
    \label{fig:hemoglobinDelay}
\end{figure}

The hemoglobin sample showed a $\sim2.5~{\rm ps}$ decay curve (top row, Fig.~\ref{fig:hemoglobinDelay}). For a discussion of the physics behind the transient absorption response of hemoglobin at these wavelengths, we refer the reader to~\cite{HemoglobinPP_1_Fu_2019}. Interestingly, hemoglobin demonstrated a pump polarization-dependent signal with differences between the polarization sensitivity of the transient absorption versus transient phase measurements (bottom row, Fig.~\ref{fig:hemoglobinDelay}). This would be impossible to observe if the pump-combining dichroic was placed before the probe-reference splitting calcite like in \cite{Orrit,Guillet_2015}. With transient absorption the pump-reference overlap showed a polarization dependence while the pump-probe overlap did not. For the pump-reference overlap, the strongest signal occurred with an s-polarized pump. With transient phase both overlaps demonstrated a polarization dependence; the pump-reference overlap shows the largest signal for the s-polarized pump and the pump-probe overlap shows the largest signal for the p-polarized pump. The polarization dependence is stronger at shorter delays (< 1 ps), similar to pump-probe anisotropy measurements on other hemeproteins~\cite{Kholodenko2000}.

\paragraph{Red Blood Cell Imaging} 

\begin{figure}
    \centering
    \includegraphics[scale = 0.83]{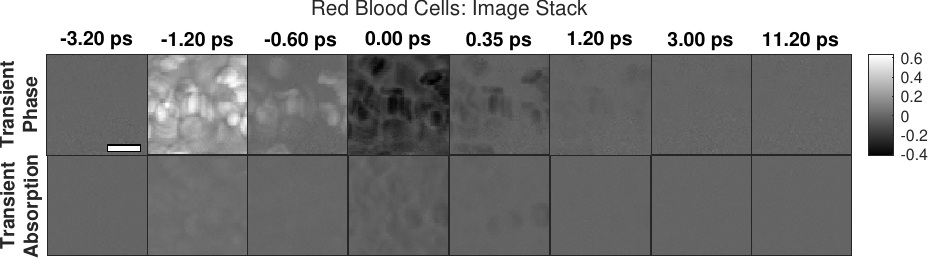}
    \caption{Image delay stack of fresh red blood cells. The scale bar is 10 $\mu$m wide; the intensity axis denotes signal (in V). The pump power at the focus is 0.76 mW and the probe power at the focus is 0.75 mW.}
    \label{fig:RBCDelayStack}
\end{figure}

Imaging human red blood cells, we expected to see a predominant signal from hemoglobin and a similar signal level from the two techniques (based on the results from the delay scans). However, T$\Phi$M demonstrated a much larger signal,  with clearer structural details (Fig.~\ref{fig:RBCDelayStack}). All of the images are shown on the same intensity scale; the overlaps with the pump-reference at $\tau=-1.20$ ps and with the pump-probe at $\tau=0$ ps show the strongest signals and good structural detail for T$\Phi$M.  

\section{Discussion}

\subsection{Impacts of Dichroic Beam-Combiner on SNR}
The pump-combining dichroic was placed \emph{after} the probe-reference splitting calcite, unlike prior works which placed the dichroic before the splitter to avoid splitting the pump~\cite{Orrit, Guillet_2015}. This enabled characterization of polarization dependence, as well as selection of relative pump-probe polarization that maximizes signal for imaging. It should be noted, however, that the dielectric coating of the dichroic is subject to polarization aberrations, including dichroism (polarization-dependent absorption), depolarization, dispersion, and retardance \cite{DichroicAberrations}. These effects can depend on both incident polarization state and angle of incidence, and can ultimately interfere with our detection scheme. One behavior we observed with our setup was a polarization-dependent spectral phase difference imparted to the probe and reference from the dichroic; the pulse broadening effects were minimal and inconsequential with our pulse durations (>300 fs), though we did notice a variation in spectral fringe visibility as a function of wavelength and the development of ellipticity on the probe and reference pulses. Additionally, our reference and probe pulses were found to have unequal amplitudes at the detectors. For TAM, only the amplitude differences contribute to a reduced SNR due to a DC offset. For T$\Phi$M, all three effects contribute to a reduced SNR. Thus, future work to address these issues will likely further the advantage of T$\Phi$M over TAM in this scenario.

\subsection{Impact of Galvanometer-scanning on Re-timing Phase in Imaging}
\begin{figure}
    \centering
    \includegraphics[scale=0.65]{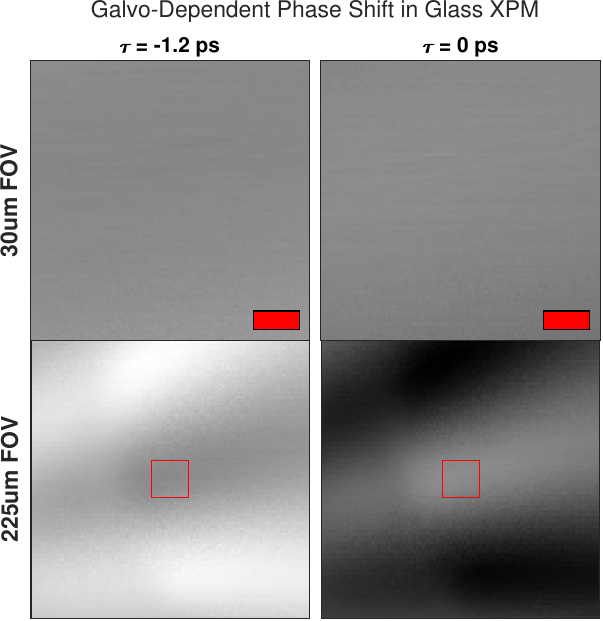}
    \caption{The top row shows the XPM images in glass measured at the pump-reference and pump-probe overlap peaks for a 30 $\mu$m x 30 $\mu$m field of view (FOV). The red scale bar is 5 $\mu$m wide. The bottom row shows the XPM images measured at the pump-reference and pump-probe overlap peaks for a 225 $\mu$m x 225 $\mu$m FOV. The square box in the center of each image outlines the FOV for the top row images. The scale for all images is the same and shows signal $\propto$ phase shift as a function of position.}
    \label{fig:galvanometerDepPhaseShift_measData}
\end{figure}

One drawback to this T$\Phi$M approach is an angle-dependent re-timing phase offset across the scanned field of view (FOV). To characterize this, an XPM image was taken in a glass slide at different FOV. Since glass is homogeneous and has no transient signal, a flat signal would be expected across the entire FOV in the absence of spatially-dependent re-timing. The measured data shown in Figure \ref{fig:galvanometerDepPhaseShift_measData} indicates a phase shift that depends on scan angle. The images shown in Figure \ref{fig:galvanometerDepPhaseShift_measData} have had the on-axis (center of image) phase shift subtracted so only the spatially dependent phase shift is portrayed.

Clearly, the larger FOV sees a stronger galvanometer-dependent phase shift $\Phi_{\rm galvanometer}$. Overall, the signal measured across a FOV will equal the sum $\Phi_{\rm galvanometer}$ + $\Phi_{\rm T \Phi \rm M}$. Decreasing the FOV or increasing the magnification of the beam at the re-timing calcite (CAL 2) could help minimize the re-timing phase effects (Supplementary 4). In order to completely eliminate this effect, de-scanning could be added in prior to CAL 2.

\subsection{Trade-offs}
One significant benefit of measuring T$\Phi$M is that XPM is measured in addition to pump-induced transients. The XPM contributes additional context about the structure of the sample related to nonlinear refractive index variations across the sample \cite{XPMSS_wilson}, though it carries no information about excited-state lifetimes. TAM will also measure instantaneous phenomena, such as two-photon absorption, but this does not give the same structural clarity as XPM and is instead specific only to materials and molecules resonant with the sum of the pump and probe energies (with peak intensities sufficient for exciting nonlinear processes).  

The microscope design presented here allows for a simple conversion between transient phase and transient absorption measurements. However, it utilizes an orthogonally polarized probe-reference pair. The probe-reference pair is required to perform the time-integrated interferometric phase measurements, but it is not required to make transient absorption measurements. Indeed, often TAM is conducted with only a pump-probe pulse pair. One downside of this design is that any sample phenomenon or optical element that spoils the polarization relationship of the probe-reference pair impacts the overall detected signal. This was discussed above for the case of the dichroic beam splitter, but the effects of birefringent samples can also be considered. 

Another potential drawback to this setup is the need for polarization purity between the probe and reference. Any loss in the ability to separate probe and reference could degrade SNR, e.g. from the dichroic, microscope optics, and the sample. Thick birefringent samples run the risk of not only changing the polarization relationship of the reference and probe pulses but also of additional pulse splitting or temporal separation. Depending on the degree to which this occurs, a very small amount of the exiting reference and probe might be within the re-timing range of the secondary calcite, or the pulses may not be able to be re-timed by the calcite at all. This would preclude the ability of this system to make phase measurements entirely.  


\section{Conclusion}

The purpose of this research was to extend T$\Phi$M, a measurement of pump-induced refractive index changes measured with interferometric probe and reference pulses, to a galvanometer-scanning configuration as a complement to TAM for  biological samples. In sum, T$\Phi$M can offer advantages over TAM, such as XPM structural contrast and potentially higher signal levels (as was the case here for hemoglobin and red blood cells), but is susceptible to degradation from polarization aberrations in the optical setup and sample birefringence, and has more limited FOV. Thus, it is beneficial to be able to acquire both TAM and T$\Phi$M with a simple rotation of a waveplate, as described here. In the future, two changes could be made to improve the utility for T$\Phi$M. Firstly, the dichroic beam combiner could be swapped for a 50/50 beamsplitter. While this would cut the power at the combiner by half, it would also eliminate the polarization-dependent effects imparted by the dichroic that we have shown negatively impact the SNR of T$\Phi$M. Secondly, de-scanning could be added prior to the re-timing calcite to remove the spatial dependence on re-timing phase. Overall, T$\Phi$M shows promise as a complementary technique to TAM, particularly for setups that may be limited in their ability to vary probe wavelengths and as a way to enhance structural context of a transient image through its simultaneous measurement of XPM.


\begin{backmatter}
\bmsection{Funding}
This research was supported by the National Institute of General Medical Sciences of the National Institutes of Health under award number R21GM135772 and by the Biophotonics Program of the National Science Foundation under award number 1943595.

\bmsection{Acknowledgments}
Research in this article was conducted at Colorado State University, Electrical \& Computer Engineering Department (CNC, RAB) and School of Biomedical Engineering (RAB). Transient absorption beamline and laser scanning microscope initially constructed by Dave Winters and Jeff Field, and modified by Patrick Stockton before adding phase detection.

\bmsection{Disclosures}
The authors declare no conflicts of interest.

\bmsection{Data Availability Statement}
Data underlying the results presented in this paper are not publicly available at this time but may be obtained from the authors upon reasonable request.

\bmsection{Supplemental document}
See Supplement 1 for supporting content. 

\end{backmatter}


\bibliography{References}

\end{document}


\maketitle

\section{System Characterization and Initialization}

When setting up the pulse splitting and re-timing calcites there are a few measurements that can be made to verify proper orientation. These measurements are described next.

\subsection{Calcite 1}

The first calcite controls two parameters: the time delay $\tau_2$ between the reference and probe pulses and the phase offset $\Phi_0$ between the reference and probe pulses. The temporal pulse separation $\tau_2$ at normal incidence is governed by the group indices for the extraordinary and ordinary axes and the thickness $d$ of the calcite:

\begin{equation}
    \tau_{2} = \frac{n_{g,o} - n_{g,e}}{c}d = \frac{1.6737 - 1.4920}{2.9979 \times 10^8 ~\text{m/s}} 2.000 ~\text{mm} \approx 1.21~\text{ps}
\end{equation}

\begin{figure}[htbp]
    \centering
    \includegraphics[scale = 0.45]{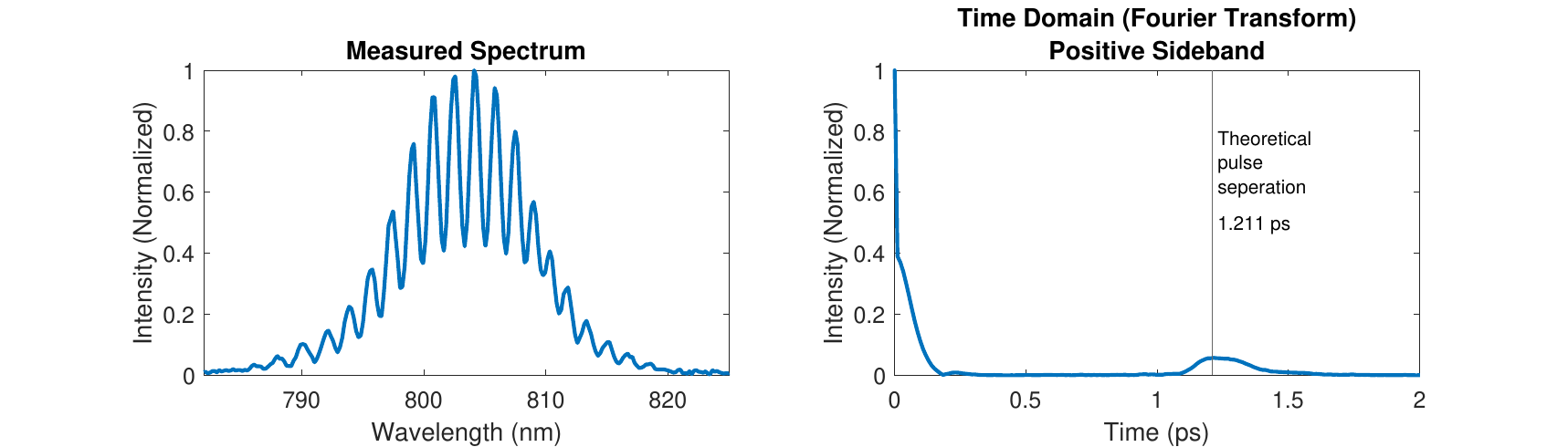}
    \caption{A Fourier transform of the spectral interference fringes measured after the first calcite reveal the expected reference-probe temporal pulse seperation of 1.2 picoseconds.}
    \label{fig:FirstCalcite_SpectralFringes}
\end{figure}

The anticipated pulse separation with a 2 mm thick piece of calcite is 1.21 ps. In order to verify this, a half-wave plate (HWP) was placed directly after CAL 1 to rotate the polarization back to 45 degrees before passing both pulses into a polarizing beam splitter. The interference spectrum was then measured at the output of the transmitted (p-polarized) light. Figure \ref{fig:FirstCalcite_SpectralFringes} shows the measured spectral fringes and the corresponding Fourier transform; as expected, the fringes correspond to a temporal separation of $\sim$1.2 ps.

\begin{figure}[htbp]
    \centering
    \includegraphics[scale=0.45]{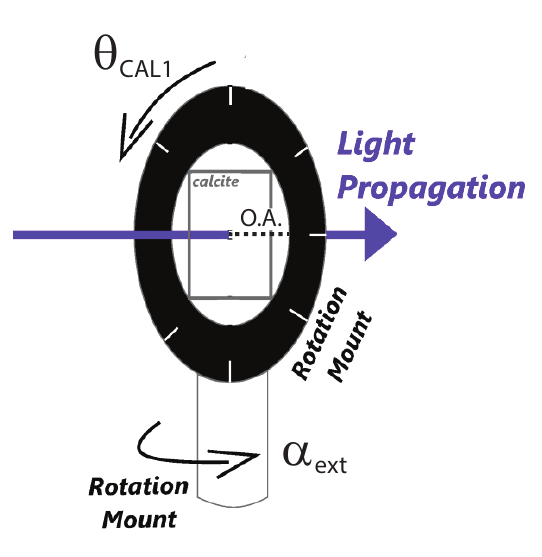}
    \caption{Diagram of calcite 1 mount. $\theta_{\rm CAL1}$ controls the orientation of the optic axis (O.A.) and $\alpha_{\rm ext}$ controls the angle of incidence which changes the offset phase $\phi_0$ between the probe and reference.}
    \label{fig:rotationMountCAL1}
\end{figure}

From the way that CAL 1 is mounted, two angles can be rotated. Figure \ref{fig:rotationMountCAL1} provides a visual of this mount. $\theta_{\rm CAL1}$ controls the relative angle of the optic axis (which is coincident with the extraordinary axis $n_e$) with the polarization of incident light. As a reminder, the polarization incident on CAL 1 has been set to 45 degrees where p-polarization is defined as parallel to the optical table. When $\theta_{\rm CAL1}$ is correctly set an equal projection of the incident pulse will be split to the extraordinary and ordinary axes. $\alpha_{\rm ext}$ controls the angle of incidence between the propagation vector of the incident light and the front plane of the calcite crystal; $\alpha_{\rm ext}$ is what changes the phase offset $\Phi_0$ between the reference and probe pulses. 

\subsection{Calcite 2}

\begin{figure}[H]
    \centering
    \includegraphics[width=1\linewidth]{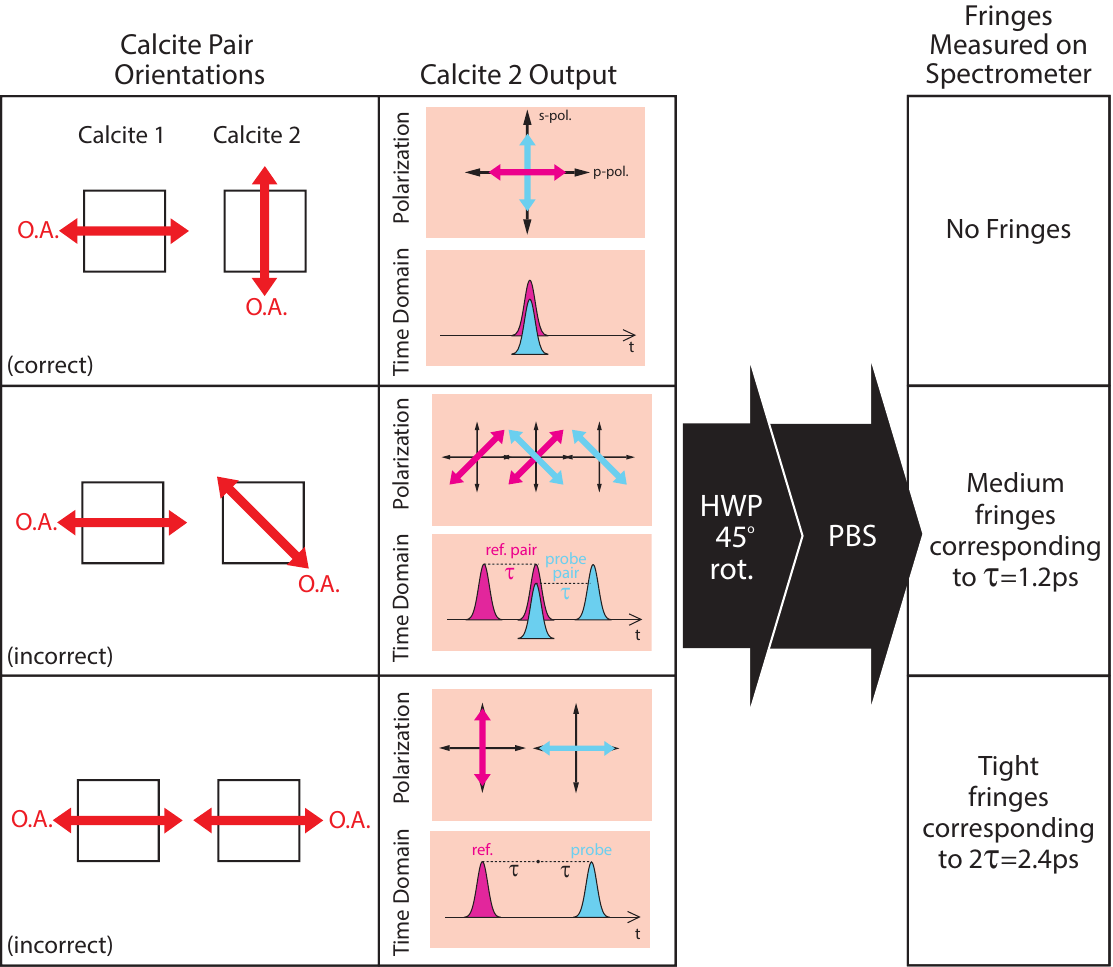}
    \caption{The effects of calcite 2 on the reference and probe pulses for different optic axis orientations. Row 1 shows the proper re-timing scenario where the optic axes (O.A.) of calcite 1 and calcite 2 are orthogonal. In this case no spectral fringes would be measured on a spectrometer. Row 2 shows the scenario with the optic axis of calcite 2 rotated 45 degrees to calcite 1; in this case each the reference and probe pulses are re-split. From the orientation of the HWP 3 though, the pair that proceeds to the spectrometer is only $\sim$1.2 ps apart. Finally, row 3 shows the scenario where the reference and probe pulses are further separated, resulting in spectral fringes $\sim $2.4 ps.}
    \label{fig:CAL2_Orientations}
\end{figure}

The second calcite is responsible for proper re-timing of the reference and probe pulses. This is critical for temporal interference to occur and thus phase measurements to be made. Similar to CAL 1, there is an angle $\theta_{\rm CAL2}$ that controls the relative alignment of the optic axis. Figure \ref{fig:CAL2_Orientations} demonstrates three possible scenarios that can occur to the reference-probe pair depending on the orientation of CAL 2.

\begin{figure}[H]
    \centering
    \includegraphics[scale = 0.75]{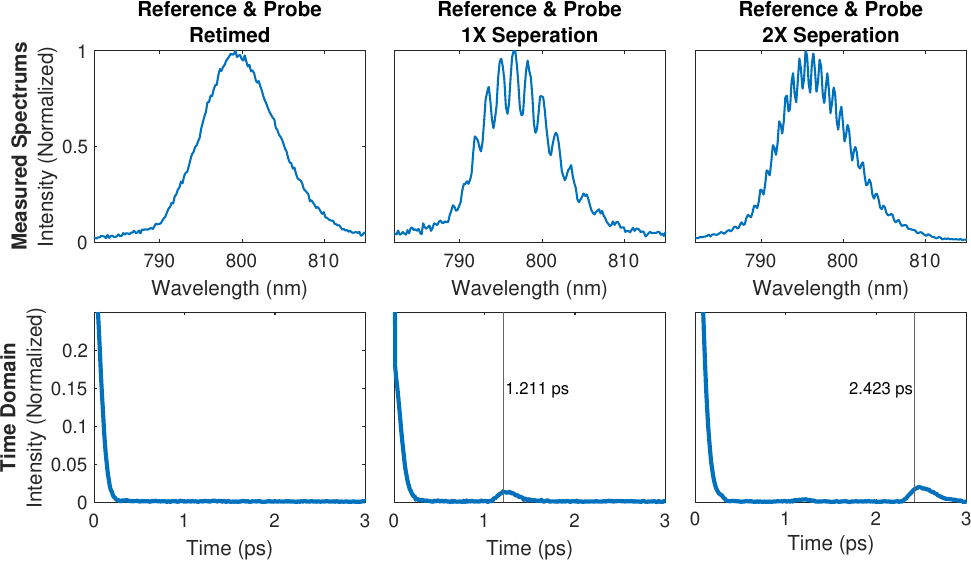}
    \caption{The interference spectra measured for different rotation angles of the re-timing calcite. The bottom row depicts the Fourier transform of each spectrum where the positive sideband is shown. Peaks can be seen for the bottom center and bottom right image corresponding to $\tau_2 = 1.2$ ps and $2\tau_2$. }
    \label{fig:BothCalcites_SpectralFringes}
\end{figure}

The ideal scenario occurs in the top row of Figure \ref{fig:CAL2_Orientations} where the optic axis of CAL 2 is perpendicular to that of CAL 1; in this case the reference and probe are properly re-timed. A confirmation of re-timing can be made by placing a spectrometer in place of one of the detectors. When the pulses are overlapped in time, no spectral fringes will be visible on the spectrometer; column 1 in Figure \ref{fig:BothCalcites_SpectralFringes} shows an example of a measured spectrum corresponding to this scenario. 

Two other undesirable scenarios can occur. One is shown in the second row of Figure \ref{fig:CAL2_Orientations} where the axes of the second calcite are rotated 45 degrees from the axes of the first crystal. The result is that both the advanced reference pulse and the delayed probe pulse are re-projected onto the slow and fast axes of the second calcite crystal splitting into two more pulses and ultimately creating four probe pulses. After HWP 3 rotates these pulses 45 degrees, the reference pulse pair is sent to one detector and the probe pulse pair to the other detector. Thus at the spectrometer spectral fringes corresponding to a 1.2 ps separation are measured; column 2 in Figure \ref{fig:BothCalcites_SpectralFringes} shows an example of a measured spectrum corresponding to this scenario. 

The final scenario that can occur is when axes of the second calcite align with the axes of the first calcite, further delaying the probe pulse and advancing the reference pulse. This results in two pulses which are separated by $2\tau_2$ manifesting as spectral fringes on the spectrometer that are twice the frequency as seen in the most recent scenario; column 3 in Figure \ref{fig:BothCalcites_SpectralFringes} shows an example of a measured spectrum corresponding to this scenario.

\section{Polarization Modeling}
The offset phase between pump and probe is given by
\begin{equation}
    \Phi_{\rm offs}=\Phi_0 + \Phi_{\rm pr,1} + \Phi_{\rm pr,2} - \Phi_{\rm ref,1} (\theta') - \Phi_{\rm ref,2},
\end{equation}
where $\Phi_0$ is a static phase offset imparted by the sample and any birefringent optics between CAL1 and CAL2, $\Phi_{\rm pr,1}$ and $\Phi_{\rm pr,2}$ are the phases on the probe pulse from CAL 1 and CAL 2, respectively, and $\Phi_{\rm ref,1} (\theta')$ and $\Phi_{\rm ref,2}$ are the phases on the reference pulse from CAL 1 and CAL 2. The total phase measured by the system will be $\Phi_{\rm total} = \Phi_{\rm offs} + \Delta \Phi_{\textnormal{t} \phi \textnormal{m}}$, where  $\Delta \Phi_{\textnormal{t} \phi \textnormal{m}}$ is the transient phase term imparted by the sample onto the probe beam in an excited state. In order to relate the phase measurements to voltage, a model is described below that connects $\Phi_{\rm total}$ to $V_{\rm A}-V_{\rm B}$.

\subsection{Model Components} 
There are several components that need to be accounted for in this model: the phase imparted by CAL 1, the phase imparted by CAL 2, and the angle of HWP 3 which controls the amplitude projection from the polarizing beam splitter onto detectors A and B.

\subsubsection{Calcite Model}

 \label{sec:R21_calciteComputationalModel}
\begin{figure}[H]
    \centering
    \includegraphics[scale = 0.7]{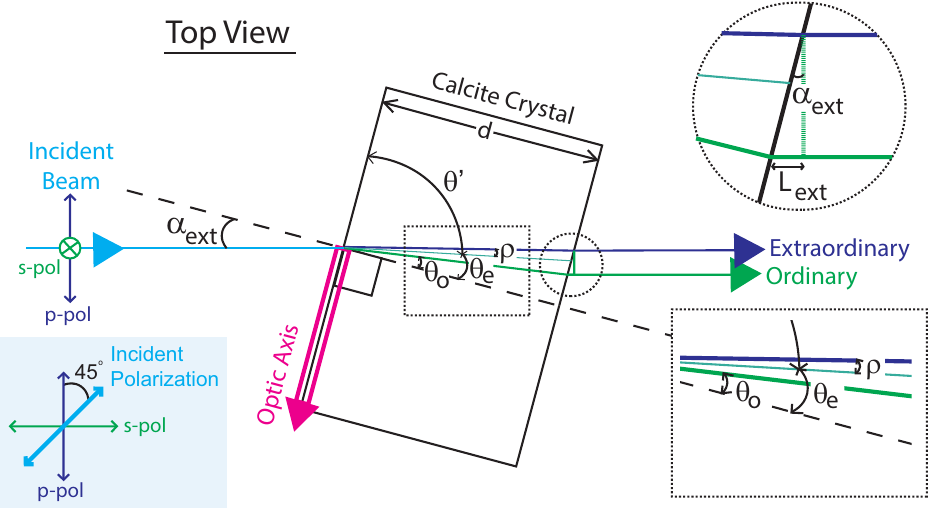}
    \caption{Diagram of the first calcite crystal (based on the diagram in ~{\cite{Schlup}} and adapted for our setup). $\alpha_{\rm ext}$ is the external angle of incidence, $\theta_0$ is the internal refraction angle for the ordinary ray, $\theta_e$ is the internal refraction angle for the extraordinary ray, $\rho$ is the walk-off angle and $\theta' = 90 - \theta_e$.}
    \label{fig:CalciteDiagram}
\end{figure}

\begin{figure}[H]
    \centering
    \includegraphics[scale = 0.65]{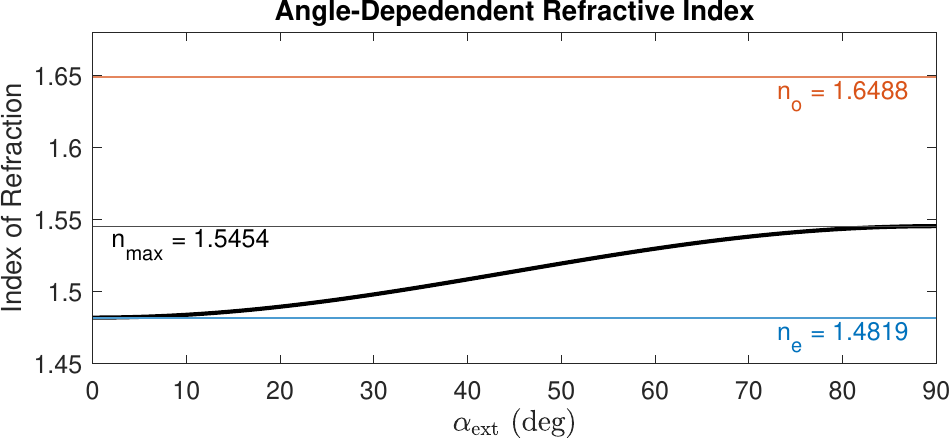}
    \caption{Angle-dependent index of refraction.}
    \label{fig:AngleDepRI}
\end{figure}

Figure \ref{fig:CalciteDiagram} shows a diagram of the calcite crystals used in this setup. The optic axis is cut along the front face of the crystal (a-cut) and creates an angle-dependent extraordinary axis index of refraction $n_e(\theta ')$. The calcite is $d = 2$ mm thick. Snell's law determines the relationship between incident and transmitted angles at an interface with $n_i \sin \theta_i  = n_t \sin \theta_t$  \cite{Peatross_Ware_PhysicalOptics}. This relationship holds simply for the ordinary ray of light which sees the ordinary refractive index. Since the extraordinary axis is angle dependent, an extra step is needed to determine the refractive index value as a function of the internal refracted angle $\theta_e$ measured relative to the optic axis by $\theta$'. Figure \ref{fig:AngleDepRI} shows the extraordinary index of refraction $n_e$ as a function of external angle of incidence $\alpha_{\rm ext}$. The range of refractive index values for the extraordinary axis is from 1.4819 to 1.5454, calculated from the dispersion coefficients in \cite{Ghosh_1999_RIValues}.

The propagation vector of the incident light is normal to the crystal face when $\alpha_{\rm ext} = 0$ and polarized at 45 degrees; the 45 degree polarization splits at the air-crystal interface into a polarization component parallel to the optic axis (p-polarized), which sees an extraordinary index of refraction, and a polarization component perpendicular to the optic axis (s-polarized), which sees the ordinary index of refraction. Calcite is a negative uniaxial crystal where $n_e - n_o < 0$ and $n_e < n_o$. This means that the polarization component with the lower refractive index (p-polarized) will travel faster through the calcite than the component with the higher refractive index (s-polarized). This results in an advanced \textbf{reference} pulse that is p-polarized and a delayed \textbf{probe} pulse that is s-polarized.

\begin{figure}[htpb]
    \centering
    \includegraphics[scale = 0.65]{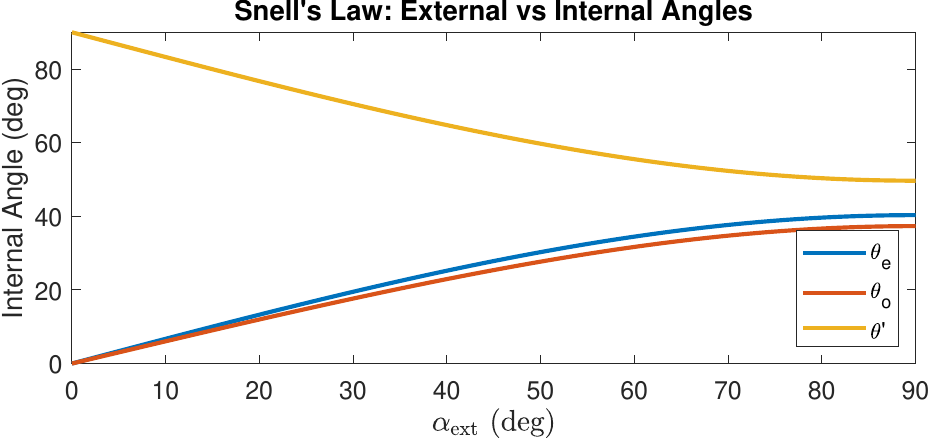}
    \caption{A plot of the internal refracted angles as a function of external angle of incidence $\alpha_{\rm ext}$}
    \label{fig:snellsLaw}
\end{figure}

Figure \ref{fig:snellsLaw} shows the internal refracted angles for the ordinary ($\theta_o$) and extraordinary ($\theta_e$) axes as a function of external angle of incidence. Theoretically, as the propagation vector rotates from perpendicular to the optic axis ($\theta$' = 90) to parallel to the optic axis ($\theta$' = 0), the extraordinary index of refraction should collapse back to the value of the ordinary index of refraction \cite{Intro2NonlinearOptics}\cite{Peatross_Ware_PhysicalOptics}. However, in our configuration since the optic axis is cut along the front face of the crystal (and we cannot set the propagation vector to lie along this axis), the maximum refracted angle $\theta_e$ that can be achieved is 40.3 degrees and the corresponding minimum angle $\theta$' is 49.7 degrees.

\paragraph{Temporal Pulse Separation}

\begin{figure}[htpb]
    \centering
    \includegraphics[scale = 0.7]{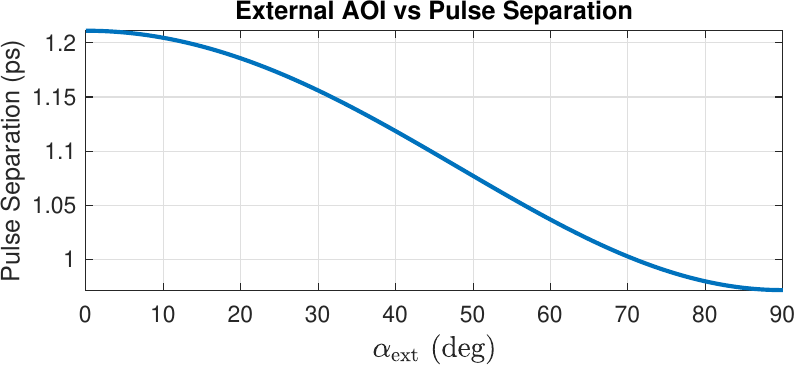}
    \caption{The simulated maximum pulse separation between the ordinary and extraordinary pulses is 1.21 picoseconds}
    \label{fig:pulseSep}
\end{figure}

The pulse separation imparted between the probe and reference pulses is a function of the external angle of incidence since the extraordinary refractive index is a function of angle. Figure \ref{fig:pulseSep} shows the calculated pulse separation as a function of $\alpha_{\rm ext}$ where the maximum pulse separation occurs at normal incidence with $\tau_2 = 1.21$ ps. 

At a wavelength of 800 nm the group refractive indices for calcite are $n_{g,o} = 1.6737$ and $n_{g,e} = 1.4920$ (calculated from the dispersion coefficients in \cite{Ghosh_1999_RIValues}); from \cite{Schlup} the equation to calculate total pulse delay is 

\begin{equation}
    T = \frac{1}{c}\left[L_{g,o} n_{g,o} + L_{\rm g,ext} - L_{g,e} n_{g,e}(\theta'_g) \right]
\end{equation}

where $c$ is the speed of light in a vacuum and $\theta'_g$ is $\theta'$ calculated using the group refractive indices. $L_{g,o} + L_{\rm g,ext}$ is the pathlength of the ordinary ray and $L_{g,e}$ is the pathlength of the extraordinary ray where the $g$ subscript indicates the use of group refractive indices. We can utilize Snell's law to find the pathlength of the ordinary ray in the calcite with

\begin{equation}
    L_{g,o} = \frac{d}{\cos (\theta_{g,o})} = \frac{n_{g,o} d}{\sqrt{n_{g,o}^2 - \sin^2(\alpha_{\rm g,ext})}}
    \label{eqn:Lo}
\end{equation}

To determine the pathlengths $L_{\rm g,ext}$ and $L_{g,e}$, we first must find $\theta_{g,e}$ and $n_{g,e}(\theta'_g)$
of the extraordinary ray (note $\theta_{g,e}$ is just $\theta_e$ calculated using the group refractive indices). Using Snell's law again and accounting for the a-cut of the crystal we can solve for $\theta_{g,e}$ by substituting Equation \ref{eqn:AngleDepRI} into Equation \ref{eqn:SnellsExtraordinary},  
\begin{equation}
    n_{g,e}(\theta'_g)\sin\theta_{g,e} = \sin \alpha_{\rm ext}
    \label{eqn:SnellsExtraordinary}
\end{equation}
where $\theta'_g = \frac{\pi}{2}- \theta_{g,e}$. The angle-dependent refractive index \cite{Schlup, Intro2NonlinearOptics, Peatross_Ware_PhysicalOptics} is found with 
\begin{equation}
    n_{g,e}(\theta'_g) = \frac{n_{g,o} n_{g,e}}{\sqrt{n_{g,e}^2  \cos^2(\theta'_g) + n_{g,o}^2 \sin^2(\theta'_g)}}
    \label{eqn:AngleDepRI}
\end{equation}
where the substitution of $\cos(\frac{\pi}{2}-\theta_{g,e}) = \sin(\theta_{g,e})$ and $\sin(\frac{\pi}{2}-\theta_{g,e}) = \cos(\theta_{g,e})$ can be utilized. Note, $\theta'_g$ in the equation above is specifically measured from the $\hat{z}$ principal axis of the calcite crystal \cite{Intro2ModernOptics}; by convention however for uniaxial crystals $n_x = n_y = n_o$ and $n_z = n_e \neq n_o$ \cite{Intro2NonlinearOptics}, meaning the principal $\hat{z}$ axis aligns with the optic axis.  After some rearranging this results in the expression
\begin{equation}
    \theta_{g,e} = \tan^{-1} \left(\frac{n_{g,o}}{\sqrt{(\frac{n_{g,o} n_{g,e}}{\sin \alpha_{ext}})^2 - n_{g,e}^2}} \right)
\end{equation}

Neglecting Poynting walkoff, we can finally find the extraordinary ray pathlength with 
\begin{equation}
    L_{g,e} = \frac{d}{\cos\theta_{g,e}}
    \label{eqn:Le}
\end{equation}
and the external ordinary pathlength
\begin{equation}
    L_{\rm g,ext} = \left[\tan(\theta_{g,e}) - \tan(\theta_{g,o}) \right] d \sin (\alpha_{\rm g,ext})
    \label{eqn:Lext}
\end{equation}

\paragraph {Phase Difference between Pulses}
In order to find the phase difference between the ordinary and extraordinary rays imparted by the first calcite we can use the same equations except with strictly the phase refractive indices $n_o$ and $n_e$ instead of the group refractive indices. This means that the equation for the angle of refraction of the extraordinary ray becomes 
\begin{equation}
    \theta_e = \tan^{-1} \left(\frac{n_{o}}{\sqrt{(\frac{n_{o} n_{e}}{\sin \alpha_{ext}})^2 - n_{e}^2}} \right)
\end{equation}
the extraordinary refractive index becomes
\begin{equation}
    n_e(\theta') = \frac{n_o n_e}{\sqrt{n_e^2  \cos^2(\theta') + n_o^2 \sin^2(\theta')}}
\end{equation}
The phase on the reference pulse from the first calcite is 
\begin{equation}
    \Phi_{\rm ref,1}(\theta ') = (\frac{L_e n_e(\theta')}{c})\omega
\end{equation}
and the phase on the probe pulse from the first calcite is 
\begin{equation}
    \Phi_{\rm pr,1} = (\frac{L_o n_o + L_{ext}}{c})\omega
\end{equation}
where  $L_{ext}$ is recalculated using the new $\theta_e$. 

\begin{figure}[H]
    \centering
    \includegraphics[scale = 0.75]{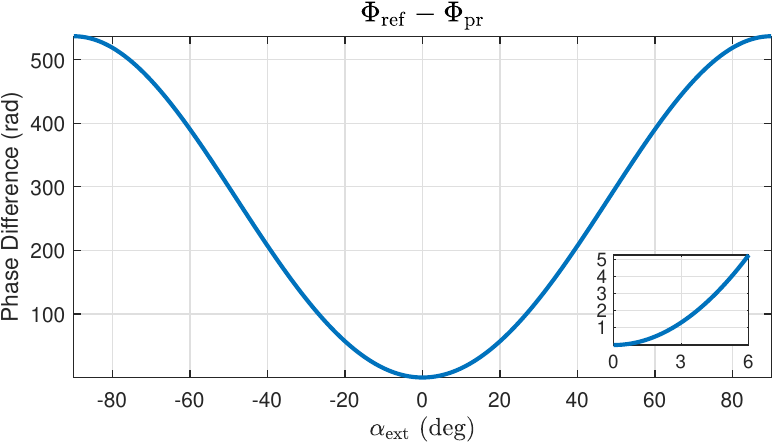}
    \caption{Total phase imparted after the second calcite crystal. The external incidence angle refers to the incidence angle on the first calcite only; the second calcite is oriented at normal incidence.}
    \label{fig:phaseSep2ndCal}
\end{figure}

At the second (re-timing) calcite the phase on the reference pulse from the ordinary axis is 
\begin{equation}
    \Phi_{\rm ref,2} =  (\frac{L_o n_o}{c})\omega 
\end{equation}
and the phase on the probe pulse from the extraordinary axis at normal incidence is 
\begin{equation}
    \Phi_{\rm pr,2} =  (\frac{L_e n_e}{c}) \omega
\end{equation}
As a reminder, the CAL 1 and CAL 2 axes are orthogonal so that the reference pulse sees the extraordinary axis for CAL 1 and the ordinary axis for CAL 2. Likewise, the probe pulse sees the ordinary axis for CAL 1 and the extraordinary axis at normal incidence for CAL 2. Thus, the total phase imparted to the reference pulse from calcite 1 and 2 is 
\begin{equation}
    \Phi_{\rm ref} = \Phi_{\rm ref,1}(\theta ') + \Phi_{\rm ref,2} = (\frac{L_e n_e(\theta')}{c})\omega + (\frac{L_o n_o}{c})\omega 
\end{equation}
 and the phase on the probe pulse is 
 \begin{equation}
    \Phi_{\rm pr} = \Phi_{\rm pr,1} + \Phi_{\rm pr,2} =  (\frac{L_o n_o + L_{ext}}{c})\omega + (\frac{L_e n_e}{c}) \omega
\end{equation}
Figure \ref{fig:phaseSep2ndCal} shows the phase difference imparted to the probe and reference pulses just from rotating the angle of incidence on the first calcite.

\subsubsection{HWP 3 Model}

\begin{figure}[H]
    \centering
    \includegraphics{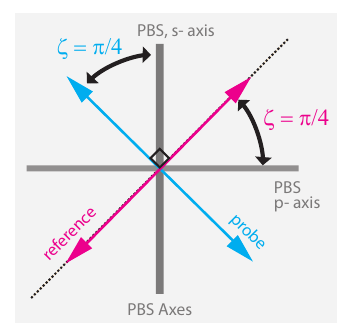}
    \caption{For the reference pulse, the cosine term of $\zeta$ proceeds to detector A and the sine term to detector B. For the probe pulse, the sine term of $\zeta$ proceeds to detector A and the cosine term to detector B.}
    \label{fig:HWP3AmplitudeProjection}
\end{figure}

HWP 3 is used to rotate the polarization axes of the reference and probe pulses after they exit calcite 2. The polarization axes are set to 45 degrees relative to the PBS axes so that both the reference and probe pulses equally project onto the s- and p- axes. Detector A measures the waveforms that transmit on the p- axis and detector B measures the waveforms that reflect on the s- axis. Figure \ref{fig:HWP3AmplitudeProjection} shows a diagram relating the polarization axes of the reference and probe pulses to the axes of the PBS. When $\zeta = \frac{\pi}{4}$ the amplitude of each pulse is split equally to either detector. If $\zeta$ is any other value, the amplitude projection of the reference pulse to the two detectors will be related by 
\begin{equation}
    E_{\rm ref}^A \propto \left(\frac{\cos(\zeta)}{\cos(\zeta)+\sin(\zeta)}\right)
\end{equation}
\begin{equation}
    E_{\rm ref}^B \propto \left(\frac{\sin(\zeta)}{\cos(\zeta)+\sin(\zeta)}\right)
\end{equation}
 and the amplitude of the probe projection will be swapped (ie. the cosine term will correspond to s- polarization and the sine term to p- polarization)
\begin{equation}
    E_{\rm pr}^A \propto \left(\frac{\sin(\zeta)}{\cos(\zeta)+\sin(\zeta)}\right)
\end{equation}
\begin{equation}
    E_{\rm pr}^B \propto \left(\frac{\cos(\zeta)}{\cos(\zeta)+\sin(\zeta)}\right)
\end{equation}
 
Where $E_{\rm ref}^A$ and $E_{\rm ref}^B$ are the electric fields incident on the A and B detectors from the reference pulse, respectively, and $E_{\rm pr}^A$ and $E_{\rm pr}^B$ are the electric fields incident on the A and B detectors from the probe pulse.

\subsection{Model Expression}
We can create a model for the balanced detection signal in the absence of the pump pulse by starting with the reference and probe waveforms estimated as Gaussian envelopes with a carrier frequency $\omega$, then adding in phase terms picked up between the first and second calcites. The reference waveforms incident  on detector A and detector B are
\begin{equation}
    E_{\rm ref}^A = Ae^{-(t^2/\sigma^2)} e^{i \omega t}
    e^{i \Phi_{\rm ref,1}(\theta ')}
    e^{i\Phi_{\rm ref,2}}
    \left(\frac{\cos(\zeta)}{\cos(\zeta)+\sin(\zeta)}\right)
\end{equation}
\begin{equation}
    E_{\rm ref}^B =  Ae^{-(t^2/\sigma^2)} e^{i \omega t}
    e^{i \Phi_{\rm ref,1}(\theta ')}
    e^{i\Phi_{\rm ref,2}}
    \left(\frac{\sin(\zeta)}{\cos(\zeta)+\sin(\zeta)}\right)
\end{equation}
and the two probe waveforms measured are 
\begin{equation}
    E_{\rm pr}^A = -A(e^{-((t-t_o)^2/\sigma^2)} e^{i \omega t} 
    e^{i \Phi_{\rm pr,1}}
    e^{i\Phi_{\rm pr,2}}
    \left(\frac{\sin(\zeta)}{\cos(\zeta)+\sin(\zeta)}\right)
    e^{i \Phi_0}
    \label{eqn:prOutOfPhase}
\end{equation}
\begin{equation}
    E_{\rm pr}^B = A e^{-((t-t_o)^2/\sigma^2)} e^{i \omega t} 
    e^{i \Phi_{\rm pr,1}}
    e^{i\Phi_{\rm pr,2}}
    \left(\frac{\cos(\zeta)}{\cos(\zeta)+\sin(\zeta)}\right)
    e^{i \Phi_0}
\end{equation}
where $\Phi_0$ is the static offset phase, $A$ is an amplitude term, and $\zeta$ measures the angle between reference-probe polarization axes and the s- and p- polarization axes of the PBS; in the ideal scenario  $\zeta= \frac{\pi}{4}$ for a phase measurement. Note the negative sign in Eqn. \ref{eqn:prOutOfPhase}; this accounts for the $\pi$ phase shift between the probe and reference pulses incident on detector A (i.e. $cos(\pi) = -1$). 

The waveforms on detectors A and B are then 
\begin{equation}
   E^A =  E_{ref}^A + E_{pr}^A
\end{equation}
\begin{equation}
   E^B =  E_{ref}^B + E_{pr}^B
\end{equation}

\begin{figure}[H]
    \centering
    \includegraphics[scale = 0.8]{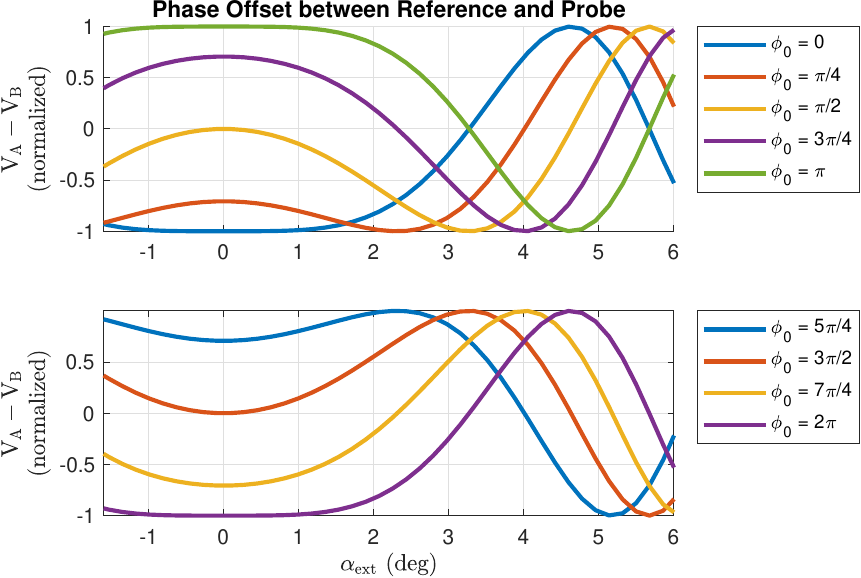}
    \caption{The impact of changing $\Phi_0$  on the signal $\Delta V$. }
    \label{fig:absolutePhaseShift}
\end{figure}

Since the detectors are slow compared with the optical carrier frequency and the pulse duration, the signal measured on each detector is the time-integrated intensity of the waveforms, 
\begin{equation}
    V_A \propto \int dt I^A = \int dt E^A E^{A*}
\end{equation}
\begin{equation}
    V_B \propto \int dt I^B= \int dt E^B E^{B*}
\end{equation}
where $V_A$ and $V_B$ are the voltage signals measured on detector A and B, respectively. Finally, for balanced detection the ultimate signal we are measuring $\Delta V_{\rm static}$ without the presence of the pump pulse is 
\begin{equation}
   \Delta V_{\rm static} = V_A - V_B \propto  -A^2\cos(\Phi_0 + \Phi_{\rm pr,1} + \Phi_{\rm pr,2} - \Phi_{\rm ref,1} (\theta') - \Phi_{\rm ref,2} )
   \label{eqn:balancedDetIdeal}
\end{equation}

where $\Delta V_{\rm static}$ is dominated by a cosine that is a function of the phase terms. The phase offset $\Phi_0$ shifts the relative location of the calibration model fringes and the associated DC offset at normal incidence as shown for a variety of phase values in Figure \ref{fig:absolutePhaseShift}. 

With the pump pulse present, we can add in our transient phase term to the final expression such that 
\begin{multline}
   \Delta V_{\rm transient} \propto -A^2\cos(\Phi_{\rm total}) \\ =  -A^2\cos(\Phi_0 + \Phi_{\rm pr,1} + \Phi_{\rm pr,2} - \Phi_{\rm ref,1} (\theta') - \Phi_{\rm ref,2} + \Delta \Phi_{\textnormal{t} \phi \textnormal{m}})
   \label{eqn:balancedDetIdeal_transient}
\end{multline}

where $\Delta V_{\rm transient}$ is the voltage measured accounting for phase offsets and the transient phase term measured through the sample after pump excitation. 

\bibliography{References_supplemental}